\documentclass{article}
\usepackage{epsfig}

\newcommand{\be}{\begin{equation}}
\newcommand{\ee}{\end{equation}}

\begin{document}

\title{Fast electromagnetic response of a thin 
film of resonant atoms with permanent dipole}

\author{J.-G.~Caputo$^{1}$ E. Kazantseva$^{1,2}$ and A.~Maimistov$^3$}
\date{\today}
\maketitle

{\normalsize \noindent $^1$:Laboratoire de Math\'ematiques, INSA de Rouen,\\
B.P. 8, 76131 Mont-Saint-Aignan cedex, France\\
E-mail: caputo@insa-rouen.fr \\
$^2$ Center for Engineering Science Advanced Research, \\
Computer Science and Mathematics Division, \\
Oak Ridge National Laboratory, \\
Oak Ridge, TN 37831, USA\\
E-mail: kazantsevaev@ornl.gov \\
$^3$: Department of Solid State Physics and Nanosystems, \\
Moscow Engineering Physics Institute,\\
Kashirskoe sh. 31, Moscow 115409, Russia \\
E-mail: amaimistov@gmail.com

}


\begin{abstract}
We consider the propagation of extremely short pulses through a dielectric
thin film containing resonant atoms (two level atoms) with permanent
dipole. Assuming that the film width is less than the field wave length,
we can solve the wave equation and reduce the problem to a system of 
generalized Bloch equations describing the resonant atoms. We compute
the stationary solutions for a constant irradiation of the film. Superimposing 
a small amplitude linear wave we compute the reflection and transmission
coefficients. From these, one can then deduce the different parameters
of the model.
We believe this technique could be used in experiments to obtain 
the medium atomic and relaxation parameters.
\end{abstract}


\section{Introduction}

One of the famous models describing field-matter
interaction is the Maxwell-Bloch system \cite{Alleb}. It
corresponds to an ensemble of two-level atoms whose states alternate
due to the electromagnetic field. In a simple
approach the electromagnetic field is assumed
to be scalar and the operator of dipole transition to only have
non-diagonal matrix elements. This model was the base for the
description of many coherent nonlinear effects, the coherent
pulse propagation (self-induced transparency \cite{McHn}) and the
coherent transient effects (optical nutations, free induction decay,
quantum beats, superradiance, photon echo). A detailed review of
these phenomena may be found in \cite{Shoe,MBES90} and in the book
\cite{Alleb}.

The Maxwell-Bloch system can be extended further than
the model of two-level atoms
discussed above. In particular one can generalize
the resonant atomic model. The model of three-level atom now attracts great
interest, because it describes electromagnetic
induced transparency, slow light propagation in (three-level) atomic
vapor~\cite{FIM05} and the coherent population transfer
\cite{Shore}. The coherent interaction of electromagnetic pulses with
quantum dots \cite{QD1,QD2} can be described by such a
generalized Maxwell-Bloch system. Another generalization of
the two-level model is to take into account
diagonal matrix elements of the dipole transition operator \cite{casperson98}.
In this medium steady state
one-half cycle pulses were obtained in the sharp line limit
\cite{mc04}. A new kind 
of steady state pulse was found,
characterized by an algebraic decay of the electric field.

The reduced
Maxwell-Bloch equation in the sharp line limit have a
zero-curvature representation\cite{Agrotis00,cm02}. A
number of results related with the complete integrability of this
reduced system have been obtained in \cite{Agrotis03a,Agrotis05,Zabolo}. The
numerical simulation of the propagation of extremely short pulses
\cite{SOE05} shows the existence of an extraordinary breather with non zero
pulse square. An analytical expression for this breather has been found in 
\cite{Agrotis00,cm02} and reproduced in \cite{Ustinov} 
using the Darboux transformation method. Recently Zabolotzki\cite{Zabolo08}
developed the inverse scattering method to find the solution for the
isotropic limit of the general model of a two-component
electromagnetic field interacting with two-level atoms with
a permanent dipole moment without invoking
the slowly varying envelope approximation.

The influence of the permanent dipole on parametric processes was
studied in \cite{Lavoine,BaBa91}. Recently Weifeng et al 
studied the generation of
attosecond pulses in a two-level system with permanent dipole 
moment\cite{Weifeng07}. For this, higher harmonics are generated
and the spectrum can be extended to 
the X-ray range. The quantum interference of both even and odd harmonics
results in the generation of higher intensity attosecond pulses.

The propagation of ultrashort pulses propagation through a thin film containing
resonant atoms located at the interface between two dielectrics is
also described by the Maxwell-Bloch equations. However,
if the width of the film is less than a wave length \cite{ry82},
the atomic system is compressed into a "point".
An insightful comment was made in
\cite{bzmt89,bmtz91}, concerning the critical role of the local Lorentz field.
This local field induces a nonlinearity so that the thin film of
two-level atoms acts as a nonlinear Fabry-Perot resonator. One 
then expects optical bistability for this device.
There are many generalizations of the thin film model,
where three (or more) level atoms, two-photon transition between
resonant levels and
non-resonant nonlinearity were studied. Here we consider a thin
film corresponding to two-level atoms with permanent dipole moment
\cite{Elyutin07a}. There the author analyzed numerically 
the pulse propagation through the film taking into
account the local field. 
He showed that a dense film
irradiated by a one-circle pulse emits a short response with a delay
much longer than the characteristic cooperative time of the
atom ensemble.
Contrary to \cite{Elyutin07a} we assume
that the atoms of the film are rare so that the local field can be neglected.

Recently \cite{ckm06} we introduced a general formalism to describe
the interaction of a (linear polarized) electromagnetic pulse with a
medium. With this we studied a ferroelectric film
which was described by a Duffing oscillator naturally giving a double well
potential. Here we generalize this approach to the case of a layer
of resonant atoms described by the generalized Bloch equations
taking into account the permanent dipole moment. An important feature
of the model is that there is a clear separation between the medium
and the surrounding vacuum so that no dispersion relation can be written.
Instead we obtain a scattering problem where the reflection
and transmission coefficients need to be calculated. \\
After presenting the model in section 2, we compute its equilibria and
their stability 
in section 3. In section 4 we assume an additional periodic
modulation around a fixed background field. This enables to do a
spectroscopy study of the film so that both the dipole parameter
and the coupling parameter can be extracted from the reflection
curve. Section 5 concludes the article.

\section{The model}

\subsection{One dimensional wave propagation}

\begin{figure}
\centerline{\epsfig{file=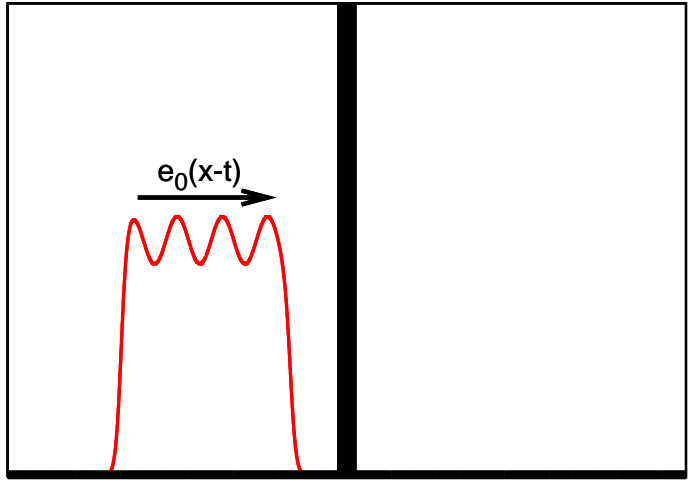,width=0.6\linewidth,angle=-90}}
\caption{Schematic of a incoming electromagnetic pulse $e_0$ on a
thin dielectric film containing resonant atoms. } \label{f1}
\end{figure}
\noindent Following the formalism of \cite{ckm06} we assume that an
electromagnetic wave is incident from the left $x<0$ on a medium
whose position is given by the function $I(x)$. The configuration
is shown in Fig. \ref{f1}.
Denoting by subscripts the partial derivatives, the equations
are
\begin{eqnarray}
e_{tt}-e_{xx} &=&g(x,t),  \label{eg} \\
g(x,t) &=&-\gamma p_{tt}I(x),
\end{eqnarray}
where $p$ is the polarization in the medium. Let the dielectric
susceptibilities of surrounding mediums be the same. That eliminates
the Fresnel refraction. The boundary conditions are \be \label{bc}
e(t, x =\pm\infty) = 0, ~~ e_t(t, x=\pm\infty) = 0, \ee and the
initial conditions are $e=e_0(x-t)$ following the scattering
problem. Then we have \be \label{Eqn4} e(t = 0, x)  = e_0(x), ~~
e_t(t = 0, x)  = -\partial_x e_0(x).
\end{equation}
We assume that the initial pulse is located at the left far from the
film.

Using the general procedure for solving the wave equation (see
\cite{ts83}),  we showed in \cite{ckm06} that the solution
of this problem is \be\label{gen_sol} e(x,t) = e_0(x-t) +
\frac{1}{2} \int\limits_{-\infty }^{+\infty } \int \limits_{0}^{t}
g(y, \tau) \left[ \theta(x-y-t+ \tau) - \theta(x-y+t - \tau) \right]
d\tau dy, \ee where we use the step-function
\[
\theta (z) = \int \limits_{-\infty }^{z} \delta (x) dx,
\]
which can also be written as
\[ \theta (z) = \left\{
\begin{array}{lr}
1 & z>0 ,\\
1/2 & z=0, \\
0 & z<0,%
\end{array}
\right.
\]

When the medium is reduced to a single thin film placed at $x=a$,
$$g(x,t) = -\gamma p_{tt} I(x),  ~~~{\rm where}~~~ I(x) = \delta(x-a). $$
The field at $x=a$ is given by
\begin{eqnarray*}
e(a,t) &=&e_{0}(a-t)+\frac{\gamma }{2}\int\limits_{0}^{t}p_{t}(a,\tau )\left[
\delta (\tau -t)-\delta (\tau -t)\right] d\tau \\
&=&e_{0}(a-t)-\gamma p_{t}(a,t).
\end{eqnarray*}
With this general formalism, one can address the question of what
happens for a medium represented by an ensemble of resonant atoms.
These could be molecules, quantum dots, or two or three level atoms.
Here we will restrict ourselves to the two-level atoms with
permanent dipole embedded into a thin film.

\subsection{The case of a thin film of resonant atoms}

\noindent Let the plane electromagnetic wave interact with the atoms
or molecules characterized by the operator of the dipole transition
between resonant energy levels and let this operator have both
non-diagonal and diagonal matrix elements \cite{casperson98}. In the
two-level approximation the Hamiltonian of the considered model can
be written as\cite{cm02}
\[
\hat{H}=\frac{\hbar \omega _{0}}{2}\left(
\begin{array}{cc}
-1 & 0 \\
0 & 1
\end{array}
\right) -\left(
\begin{array}{cc}
d_{11}E & d_{12}E \\
d_{21}E & d_{22}E
\end{array}
\right) ,
\]
where $E$ is the amplitude of the electric field of the electromagnetic
wave and $\omega_{0}$ is the frequency difference between the levels. The 
polarization of the medium is
\be\label{film_polar}
P(x,y,t)=l \delta(x-a)n_{A}p(t),\ee
where $l$ is the film thickness and where $n_{A}$ is
the volume density of the atoms. We assumed a homogeneous film
and neglected the transverse $y$ dependence of $P$.
The atomic polarizability $p$ is given by
the expression
\be \label{eqPolar} p={\rm tr} \hat{\rho}\hat{d}=\rho _{11}d_{11}+\rho
_{22}d_{22}+\rho _{12}d_{21}+\rho _{21}d_{12} \ee
$$=\frac{1}{2}\left( d_{11}+d_{22}\right) +\frac{1}{2}\left(
d_{11}-d_{22}\right) (\rho _{11}-\rho _{22})+\rho _{12}d_{21}+\rho
_{21}d_{12}, $$
where $\hat \rho$ is the
density matrix and where we used the constraint $\rho _{11}+\rho _{22}=1$.
In the expression above, the first term corresponds to the
constant polarizability of the molecules. The average of this
quantity
over all atoms must be zero because we assume no polarization
of the medium in the absence of an electromagnetic field.
Note that all relaxation processes are neglected because we assume
short electromagnetic pulses. We also neglect the dipole-dipole interaction,
ie we assume the atoms carrying dipoles are rare in the film.
The evolution of the elements of $\hat{\rho}$\ is given by the Heisenberg
equation $ i\hbar \partial \hat{\rho}/\partial
t=\hat{H}\hat{\rho}-\hat{\rho}\hat{H}$\ and yields the Bloch
equations.

So the total system of equations describing the interaction
of an electromagnetic wave with a collection of identical atoms
in a film placed at $x=a$ is \cite{casperson98}
\begin{equation}
\frac{\partial ^{2}E}{\partial x^{2}}-\frac{1}{c^{2}}\frac{\partial
^{2}E}{
\partial t^{2}}=\frac{4\pi n_{A}}{c^{2}}l \delta(x-a) \frac{\partial ^{2}}{\partial t^{2}}
\left( \frac{1}{2}\left( d_{22}-d_{11}\right)
r_{3}+d_{12}r_{1}\right),  \label{eqm}
\end{equation}
\begin{equation}
\frac{\partial r_{1}}{\partial t}=-\left[ \omega _{0}+\left(
d_{11}-d_{22}\right) E/\hbar \right] r_{2},  \label{bl1}
\end{equation}
\begin{equation}
\frac{\partial r_{2}}{\partial t}=\left[ \omega _{0}+\left(
d_{11}-d_{22}\right) E/\hbar \right] r_{1}+2(d_{12}E/\hbar )r_{3},
\label{bl2}
\end{equation}
\begin{equation}
\frac{\partial r_{3}}{\partial t}=-2(d_{12}E/\hbar )r_{2},  \label{bl3}
\end{equation}
where $r_{1,2,3}$ are the components of the Bloch vector
\begin{equation}
r_{1}=\rho _{12}+\rho _{21},\quad r_{2}=-i(\rho _{12}-\rho _{21}),\quad
r_{3}=\rho _{22}-\rho _{11}.  \label{eqBloch}
\end{equation}
In equations (\ref{bl1})-(\ref {bl3}) the field $E$ is taken in the film,
i.e., at $x=a$. This system differs from the well-known
Maxwell-Bloch equations \cite{R3,R4} by the terms containing the
parameter $\left( d_{11}-d_{22}\right) $.

We normalize time, space and electric field as
\[
\tau =\omega _{0}t,\quad \zeta =\omega _{0}x/c, \quad e=2d_{12}E/\hbar \omega
_{0}.
\]
We also introduce the parameter $\mu $ which measures the strength
of the permanent dipole:
\begin{equation}
\mu = { d_{11}-d_{22}\over 2d_{12} } .
\end{equation}
Then the Maxwell-Bloch equations (\ref{eqm})-(\ref {bl3}) take the
form:
\begin{equation}
r_{1,\tau }=-(1+\mu e)r_{2},\quad r_{2,\tau }=(1+\mu
e)r_{1}+er_{3},\quad r_{3,\tau }=-er_{2},  \label{eq27}
\end{equation}
\begin{equation}
e_{,\zeta \zeta }-e_{,\tau \tau }=\alpha (r_{1}-\mu r_{3})_{,\tau
\tau }\delta(\zeta-\zeta_0), \label{eq28}
\end{equation}
where $\zeta_0 = \omega_0 a/c$, and
\begin{equation}
\alpha =8\pi n_{A}l{\omega \over c} {d_{12}^{2} \over \hbar \omega _{0}}.
\end{equation}
A rough estimation of this parameter taking $d_{12}\approx $
1 Debye $= 3.336 ~10^{-30}$ gives 
$\alpha = n_{A}l 0.00795$ so that for a film thickness of $l=1 \mu m$  and
a volume density  $ 10^7 < n_{A} < 10^9$ we get $0.1 < \alpha < 10$. If the film
is thinner, $l=1  nm$ we can take the same range of $\alpha$ with
densities $ 10^{10} < n_{A} < 10^{12}$.  This is still compatible with
our hypothesis of dilute impurities in the medium so that we can
neglect the dipole-dipole interaction.

The system can be further simplified by noting from equation (\ref{eq27}) 
that $(r_1-\mu r_3)_\tau =-r_2$. We then get
\begin{equation}
e_{,\tau \tau }-e_{,\zeta \zeta }=\alpha r_{2,\tau
}\delta(\zeta-\zeta_0),  \label{eqm2}
\end{equation}
\begin{equation}
r_{1,\tau }=-(1+\mu e)r_{2},\quad r_{2,\tau }=(1+\mu
e)r_{1}+er_{3},\quad r_{3,\tau }=-er_{2}.  \label{eq27a}
\end{equation}
This is the model that we will analyze in detail in this
article.

We will assume the general initial conditions where the
medium is initially at rest so that $e=0,$ $e_{,\zeta }=e_{,\tau
}=0,r_{1}=r_{2}=0,$ $r_{3}=-1$, at $\tau \longrightarrow -\infty $.
From the Bloch equations (\ref{eq27}) we obtain $\left(
r_{1}^{2}+r_{2}^{2}+r_{3}^{2}\right) _{,\tau }=0.$\ and using the initial
conditions we get the value of this integral of motion
\begin{equation}
r_{1}^{2}+r_{2}^{2}+r_{3}^{2}=1.  \label{eq210}
\end{equation}
We consider the effect of an electromagnetic pulse impinging
on the thin film placed at $\zeta_0=0$. For that we can use the general result (\ref{gen_sol}) to solve the wave
equation (\ref{eqm2}). Taking the right part of (\ref{eqm2}) as
a function $g(y,\tau)$ under the integral in (\ref{gen_sol}) we obtain the
following expression

\begin{equation}
e(0,\tau )=e_{0}(-\tau )+\alpha \int\limits_{0}^{\tau }r_{2}(\tau ^{\prime
})\delta (\tau -\tau ^{\prime })d\tau ^{\prime }=e_{0}(-\tau )+\frac{\alpha
}{2}r_{2}(\tau ).  \label{ein}
\end{equation}
which represents the strength of the electrical field in the film. Now,
using this expression we can write the modified Bloch equations as
\begin{eqnarray}
r_{1.\tau } &=&-(1+\mu e_{0})r_{2}-(\alpha \mu /2)r_{2}^{2},  \label{eq61} \\
r_{2,\tau } &=&(1+\mu e_{0})r_{1}+e_{0}r_{3}+(\alpha \mu
/2)r_{1}r_{2}
+(\alpha /2)r_{2}r_{3},  \label{eq62} \\
r_{3,\tau } &=&-e_{0}r_{2}-(\alpha /2)r_{2}^{2}.  \label{eq63}
\end{eqnarray}
These are the correct equations describing the resonant responses of
two-level atoms of a thin film to an ultra-short electromagnetic
pulse. It is worth noting that the electromagnetic wave is incident
normally on the film. Second, all atoms of the film are identical.
Finally note that we did not assume any limitation on the time
duration of the electromagnetic pulse. It may be a half period
pulse, i.e. an electromagnetic spike or a quasiharmonic wave.

Finally note that since the motion occurs on the Bloch sphere,
the system (\ref{eq63}) has the constraint $r_1^2+r_2^2+r_3^2=1$.
It is then natural to write it in the reduced coordinates $(m,\phi)$
such that
\be\label{rtomphi}
r_1 = \sqrt{1-m^2}\cos \phi,~~~~r_2 = \sqrt{1-m^2}\sin \phi,r_3=m.\ee
The system is then
\begin{eqnarray}\label{mphidot}
m_\tau = -e_0 \sqrt{1-m^2}\sin \phi -{\alpha\over 2}(1-m^2)\sin^2 \phi ,\\
\phi_\tau = (\mu + {m\over \sqrt{1-m^2}}\cos \phi)
(e_0+{\alpha\over 2}\sqrt{1-m^2}\sin \phi) +1.
\end{eqnarray}

\section{Equilibrium states}

We now study the stationary points of the system (\ref{eq63}). Initially
before the wave reaches it, the film is at rest. The electromagnetic
field $e_0$ shifts the film state to a new equilibrium. The system then
relaxes back to its original state after the wave has passed.
We will examine these new transient states and their stability.

\begin{figure}
\centerline{
\epsfig{file=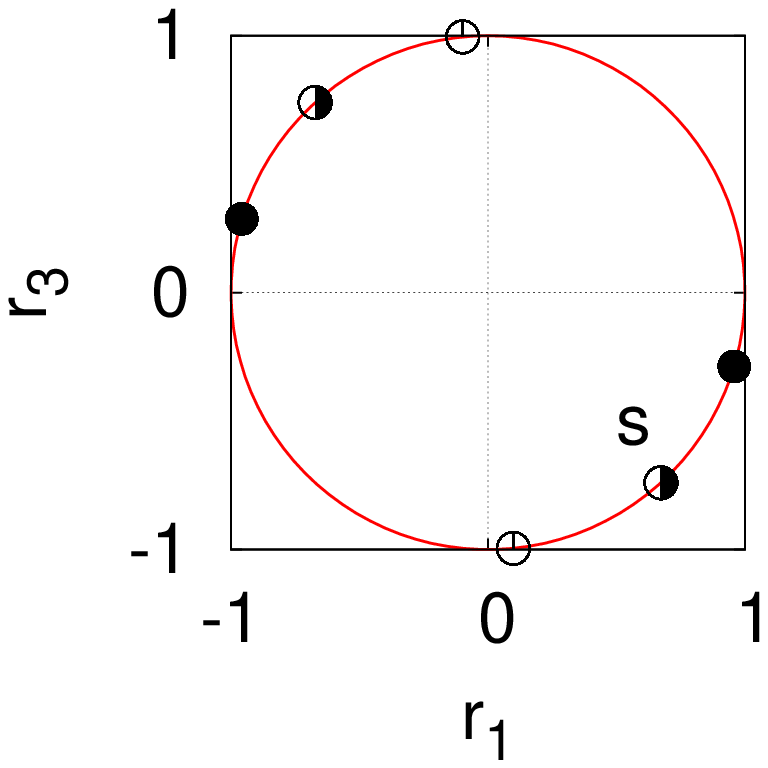,width=0.4\linewidth,angle=0}
\epsfig{file=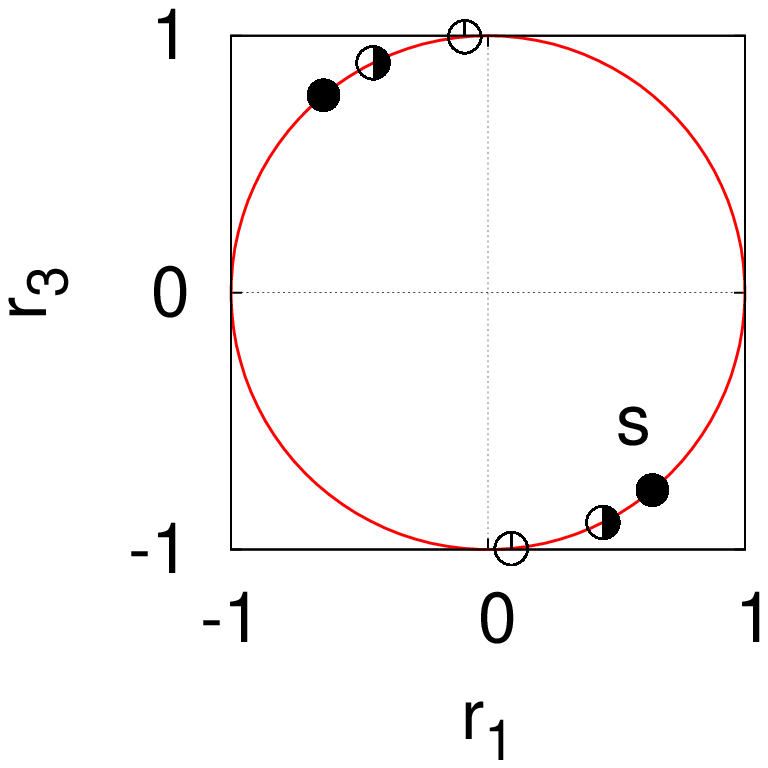,width=0.4\linewidth,angle=0}}
\centerline{
\epsfig{file=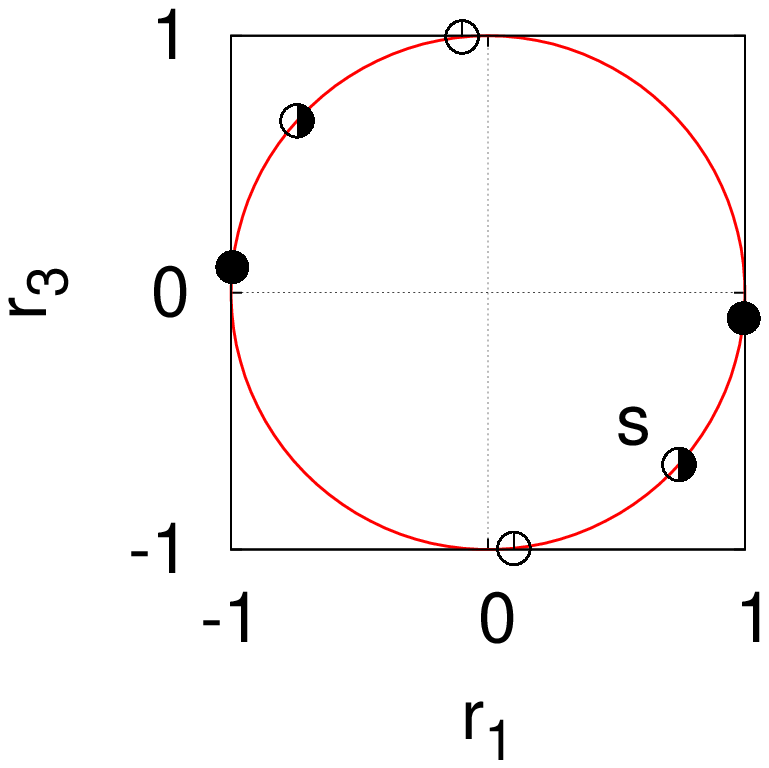,width=0.4\linewidth,angle=0}
\epsfig{file=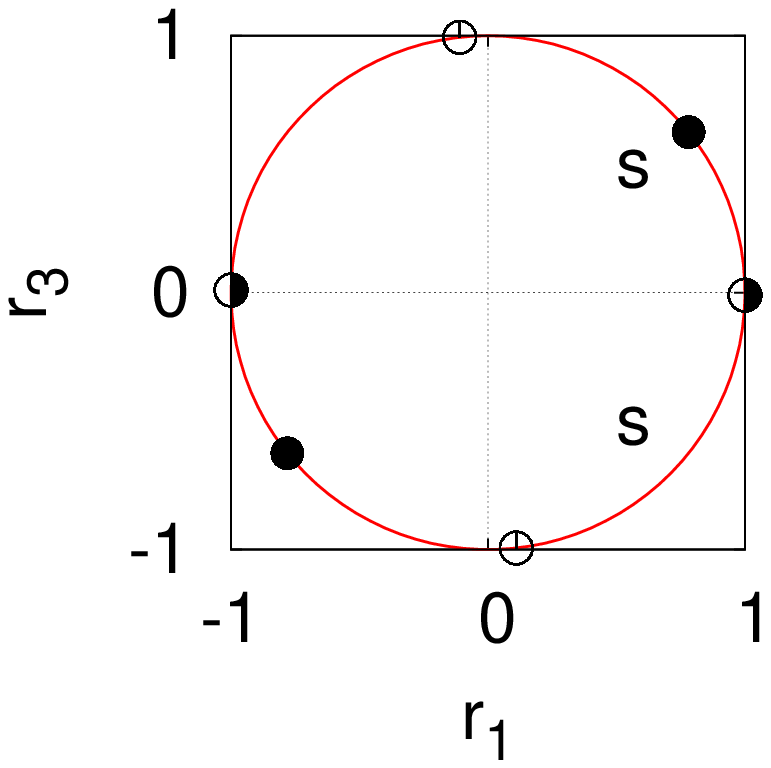,width=0.4\linewidth,angle=0}}
\caption{Position of the fixed points $(r_1^{\ast}, 0,r_3^{\ast})$
on the Bloch sphere for four different values of the dipole parameter
$\mu$, top from left to right $\mu=0.1,~1$ and bottom 
from left to right $\mu=-0.1,~-1$. The shades correspond
to different amplitudes of the electric field $e_0$, empty is
for $e_0=0.1$ , half-empty is for $e_0=1$ and full is for $e_0=5$.
The stable fixed points (see text) are indicated by the letter s. }
\label{f2}
\end{figure}

For $e_{0}$ constant, the system of
equations (\ref{eq61}-\ref{eq63}) has the fixed point
$(r_1^{\ast}, 0,r_3^{\ast})$ where  $r_{1,3}^{\ast} $ satisfy
\begin{equation}
(1+\mu e_{0})r_1^{\ast} +e_{0} r_3^{\ast}=0.  \label{eq71}
\end{equation}
Assuming $r_2^{\ast } \neq 0$ leads to a contradiction.
From (\ref{eq210}) it follows that
\begin{equation}
r_{1}^{\ast 2}+r_{3}^{\ast 2}=1  \label{eq72}
\end{equation}
Combining these two equations we obtain
\be\label{fix_point}
r_3^{\ast}= \pm [1 + {e_0^2 \over (1+\mu e_0)^2} ]^{-1/2},~~
r_1^{\ast}= -{ e_0 \over 1+\mu e_0} r_3^{\ast}.\ee
This fixed point corresponds to a stationary polarization and
population induced in the two-level atoms by the
incident constant field $e_{0}$.
When $e_{0}\rightarrow \infty $
\begin{equation}
r_{3}^{\ast }\rightarrow \pm {\mu \over \sqrt{1+\mu^2}},~~~
r_{1}^{\ast }\rightarrow - {r_{3}^{\ast } \over \mu}.
\label{FT}
\end{equation}
To understand geometrically the position of the fixed points, we can
parametrize the Bloch sphere with $r_2^{\ast }=0$ by
$$ r_1^{\ast } = \sin \theta, ~~ r_{3}^{\ast }= \cos \theta, $$
yielding the relation
\be\label{theta_fp}
\tan \theta = - {e_0 \over 1+\mu e_0}.\ee
The dependence of the values $r_{1,3}^{\ast }$ \ on the electric field $e_0$
is monotonic so there is no bistability.
The two fixed points are shown in Fig. \ref{f2} for $\mu=\pm 0.1, ~\pm 1$.
For $\mu$ small (left panel) the fixed points
start from $(0,\pm 1)$ and rotate following $(\sin\theta,\cos \theta)$
with $\theta$ given by (\ref{theta_fp}). For $\mu=-1$ and $e_0= 1$ 
$\theta= \pi/2 + n \pi$ so we obtain $(\pm 1,0)$. Since $r_3$ is the difference in the
population of the levels, these are equally populated. When $\mu$ is large
and positive (top right panel) the fixed points do not change very much when
the electric field is increased. On the contrary when $\mu $ is large and
negative (bottom right panel) increasing the electric field shifts the fixed points
from $(0,\pm 1)$ to $(\pm 1,0)$ for $e_0=1$ and about
$(\pm 1/\sqrt{2},\pm 1/\sqrt{2})$ for $e_0=5$. For this last value of the
parameter we can have a strong population inversion.

To study the stability of the fixed point we linearize
the modified Bloch equations. We set
$r_{1}=r_{1}^{\ast }+\delta r_1$, $r_{2}= 0 + \delta r_2 $,
$r_{3}=r_{3}^{\ast }+ \delta r_3$. The
linearized modified Bloch equations read
\begin{eqnarray}
{\delta r_1}_{\tau } &=&-(1+\mu e_{0})\delta r_2, \nonumber \\
{\delta r_2}_{\tau } &=&(1+\mu e_{0})\delta r_1 +e_{0}\delta r_3
+(\alpha \mu /2)r_{1}^{\ast }\delta r_2
+(\alpha /2)r_{3}^{\ast }\delta r_2,  \label{linBlo} \\
{\delta r_{3}}_{\tau } &=&- e_{0}\delta r_{2},  \nonumber
\end{eqnarray}
where as usual the $\tau$ subscript indicates the derivative.
We introduce the field modified frequency 
\be \label{big_omega}
\Omega^2 =(1+\mu e_{0})^2 +e_{0}^{2},\ee
and ratio
\be\label{defb}
b={\alpha \over 4} (\mu r_{1}^{\ast } +r_{3}^{\ast }) = 
{\alpha \over 4} {r_{3}^{\ast } \over 1 +\mu e_0} =
\pm {\alpha \over 4 \Omega}.\ee
From (\ref {linBlo}) we can get the equation for $\delta r_{2}$
\be\label{dr2}
{\delta r_{2}}_{\tau \tau }-2b {\delta r_{2}}_{,\tau }+\Omega ^{2}\delta r_{2}=0.
\ee
The characteristic equation is
\[
\lambda ^{2}-2b\lambda +\Omega ^{2}=0,
\]
whose roots are%
\[
\lambda _{1,2}=b\pm i\sqrt{\Omega ^{2}-b^{2}} = \pm {\alpha \over 4 \Omega}
\pm i \Omega \sqrt{ 1 - {\alpha^2 \over 16 \Omega^4 }}
\]
Then if $b<0$\ and $\Omega ^{2}>b^{2}$ the fixed point is stable.
Depending on $\alpha$ two cases occur. Consider first
a small $\alpha$ so that
$\Omega ^{2}>b^{2}$. Then if $\mu >0$, $b>0$ if $r_{3}^{\ast } >0$.
Then the fixed point such that $r_{3}^{\ast } >0$ is unstable.
Conversely the fixed point such that $r_{3}^{\ast } <0$ is stable.
If $1 +\mu e_0 <0 $ which occurs for $\mu <0$ and large
fields $e_0$ the situation is reversed.
The fixed point such that $r_{3}^{\ast } >0$ is stable while
the one such that $r_{3}^{\ast } <0$ is unstable. 
The oscillation frequency of the orbit as it approaches the
fixed point is given by the imaginary part of $\lambda$
\be\label{osc_freq}
\omega^{\ast } = \Omega \sqrt{1 - {\alpha^2 \over 16 \Omega^4 }}\ee.

When  $\Omega ^{2}>b^{2}$ corresponding to a large $\alpha$, there
is no imaginary part of $\lambda$. Even for very large $\alpha$ 
the stability remains unchanged because for the first term $b ={\alpha \over 4 \Omega} $ will
always dominate the second one 
$\Omega \sqrt{ {\alpha^2 \over 16 \Omega^4 }-1 }$. Therefore
the fixed point such that $r_{3}^{\ast } <0$ is stable (resp. unstable) 
for $\mu >0$ (resp. $1 +\mu e_0 <0 $). In this case
there are no oscillations around the fixed point. 

\begin{figure}
\centerline{
\epsfig{file=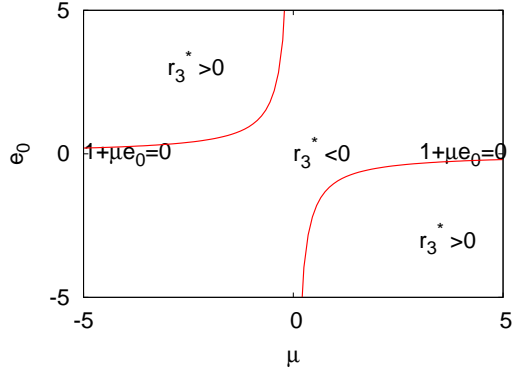,width=0.6\linewidth,angle=0}}
\caption{Stability diagram in the $(\mu,e_0)$ plane indicating 
which fixed point is stable.}
\label{f3}
\end{figure}

To conclude, for all values of $\alpha$,  the stability
is shown in Fig. \ref{f3} in the parameter space $(\mu,e_0)$.

\subsection{Numerical results}

\begin{figure}
\centerline{ 
\epsfig{file=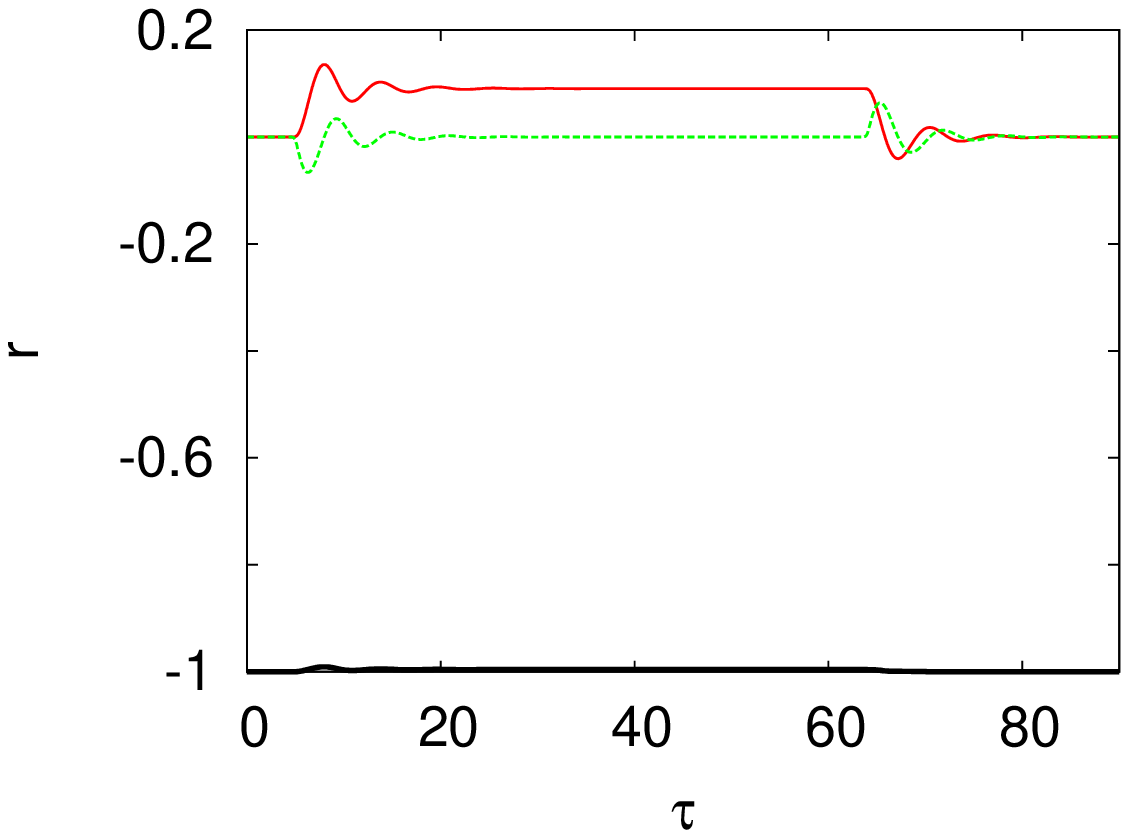,height=5 cm,width=5cm,angle=0}
\epsfig{file=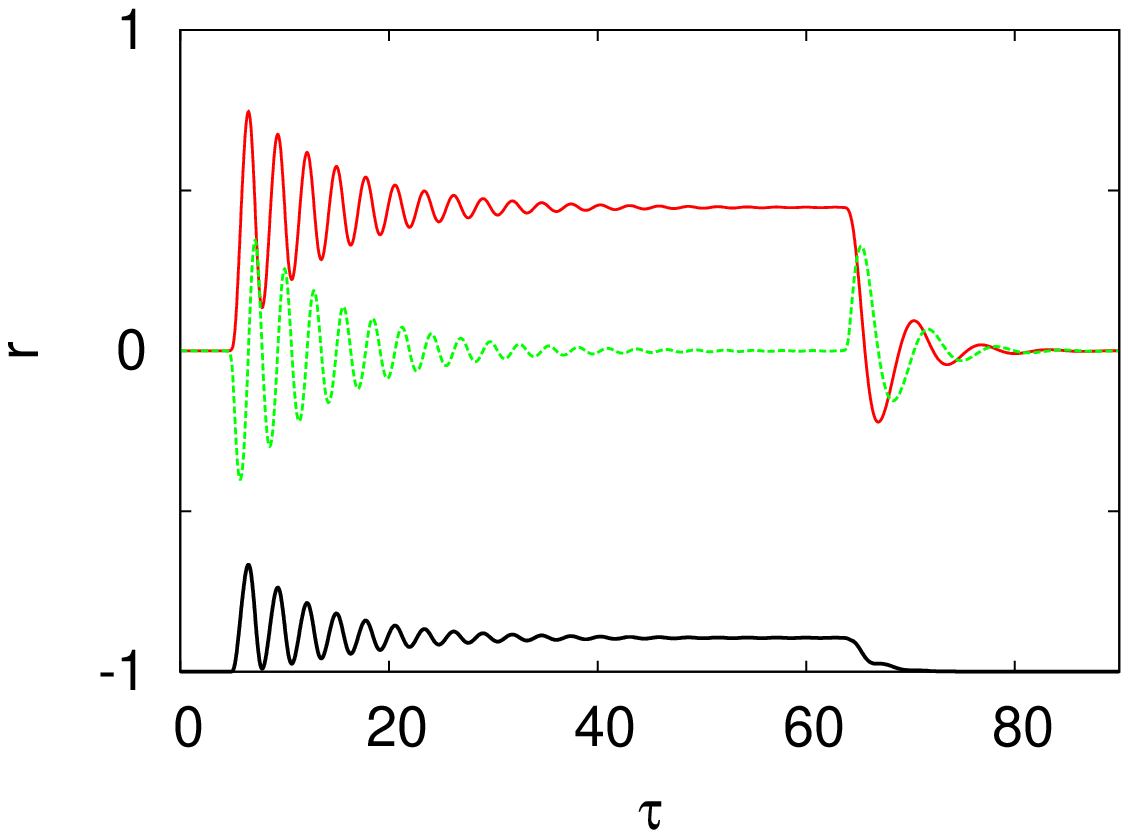,height=5 cm,width=5cm,angle=0}
\epsfig{file=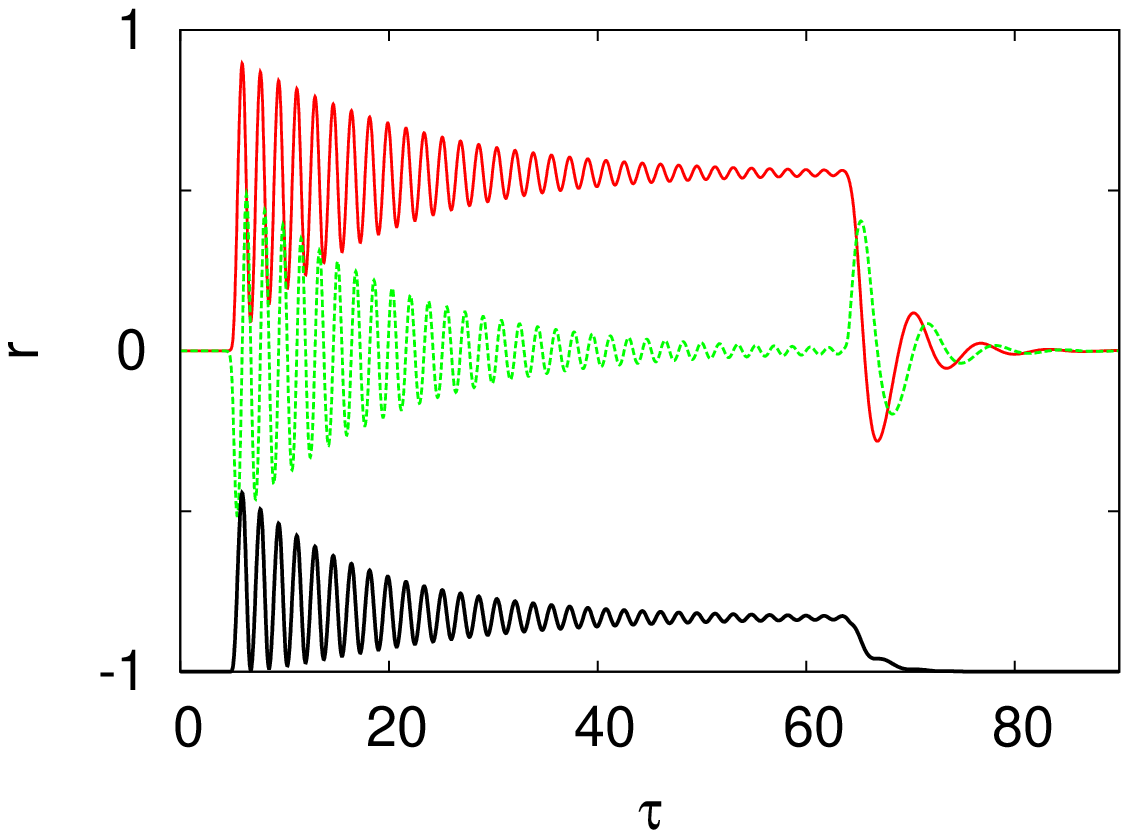,height=5 cm,width=5cm,angle=0}
}
\caption{Time evolution of the components of the Bloch vector
$(r_1,r_2,r_3)$ respectively in continuous line, long dash and short dash
(red, green and black online). From the left panel to the right panel the amplitude
of the field is increased from $e_0=0.1$ (left), $e_0=1$ (middle) to $e_0=2$ (right).
The dipole parameter is $\mu=1$ and the coupling coefficient is $\alpha=1$.
}
\label{f4}
\end{figure}

To illustrate the previous analysis, we have solved numerically the
system of ordinary differential equations (\ref{eq61}-\ref{eq63}). 
We used the Runge-Kutta 4-5 Dopri5\cite{hairer} as solver.
For all the runs presented, we choose an incoming pulse
\be\label{e0num}
e_0(-t) = e_{00} {1 \over 2} [ \tanh ({t-t_1 \over w})
-\tanh ({t-t_2 \over w})],\ee
where $t_1 = -64$, $t_2 = -5$ and $w=0.2$. In the
following we will abusively name the amplitude $e_{00}$ 
$e_0$. We first consider
$\mu=1$. Fig. \ref{f4} shows the evolution of the Bloch vector 
components as a function of time, with from left to right $e_0=0.1, 1$
and 2. As expected the system reaches the stable equilibrium
state given by (\ref{fix_point}) with $r_3^*<0$. The oscillations
are given by the frequency $\omega^*  \approx  1.1, 2.2$ and 3.6
from left to right. These plots correspond to the upper right
panel of Fig. \ref{f2}.

\begin{figure}
\centerline{ 
\epsfig{file=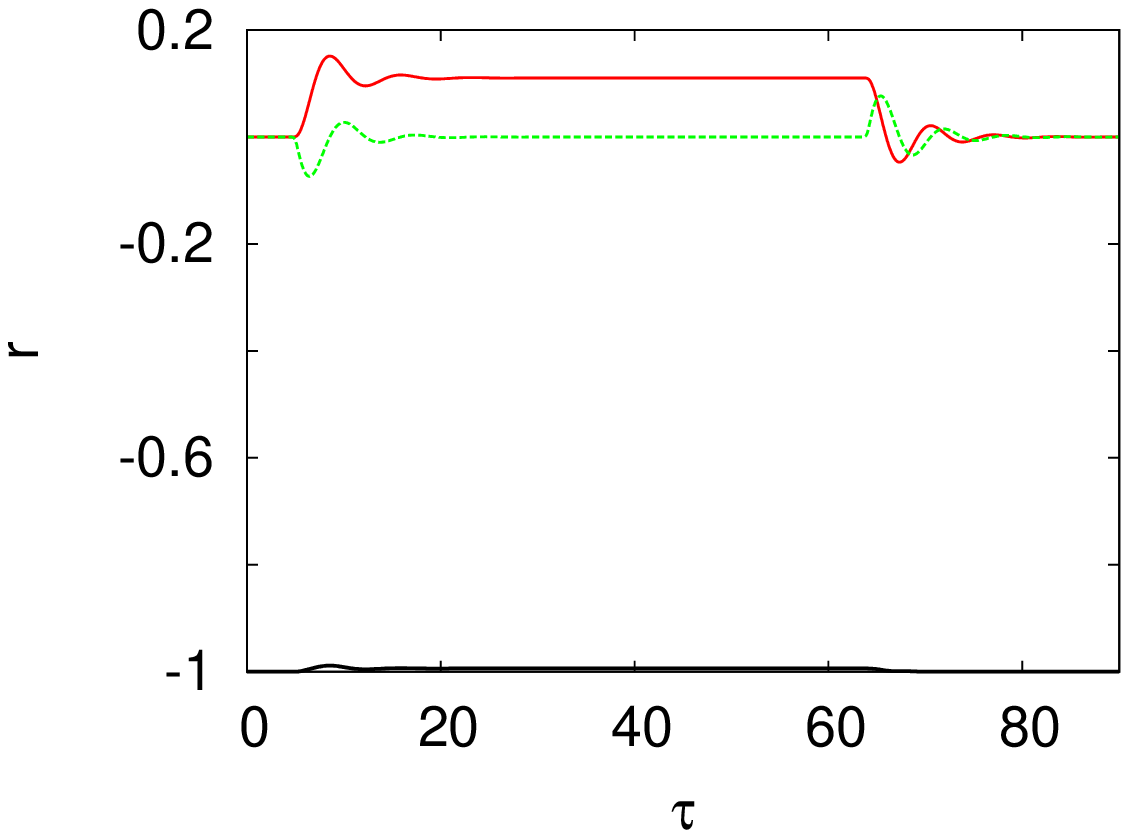,height=5 cm,width=5cm,angle=0}
\epsfig{file=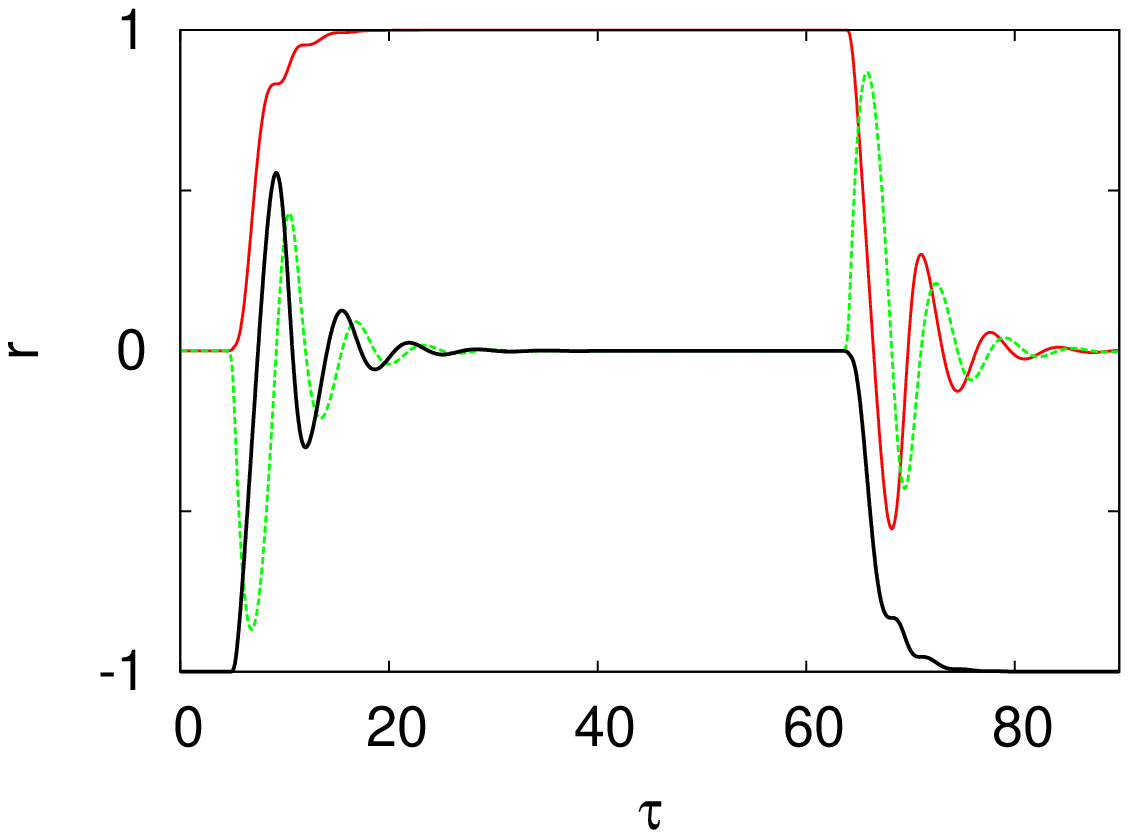,height=5 cm,width=5cm,angle=0}
\epsfig{file=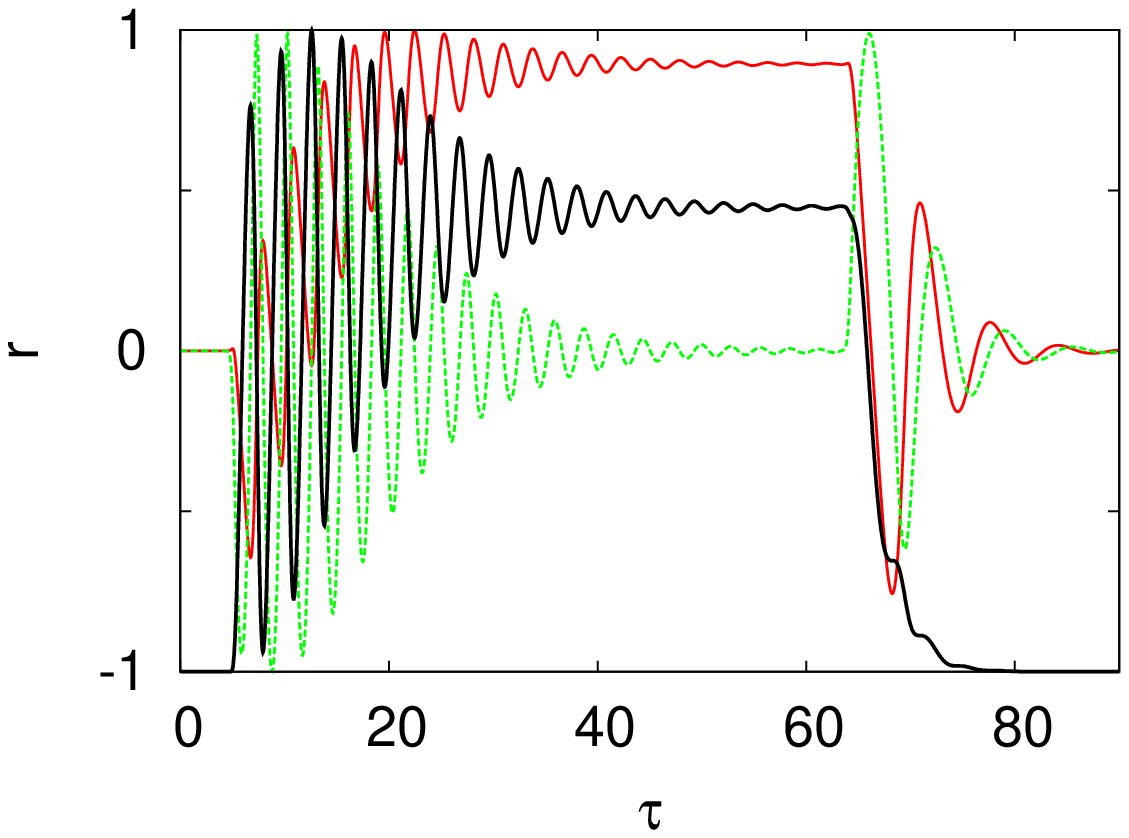,height=5 cm,width=5cm,angle=0}
}
\caption{Same as for Fig. \ref{f4} except that the dipole parameter
is $\mu=-1$.
}
\label{f5}
\end{figure}

The case $\mu = -1$ is shown in Fig. \ref{f5}. When $e_0$ is small,
we have an equilibrium very close to the one for $\mu=1$ as
expected from the bottom right panel of Fig. \ref{f2}. When
$e_0=1$ we obtain the situation where $r_3^*=0$ and $r_1^* = \pm 1$.
Note the typical frequency $\omega^*=1$ as in the left panel.
For a larger field $e_0=2$, shown in the right panel we obtain as expected
an equilibrium $r_3^* >0$. Note the relatively small oscillation
frequency $\omega^* \approx 1.7$ compared for the one for
the same $e_0$ and $\mu=1$ (right panel of Fig. \ref{f4}).

\begin{figure}
\centerline{
\epsfig{file=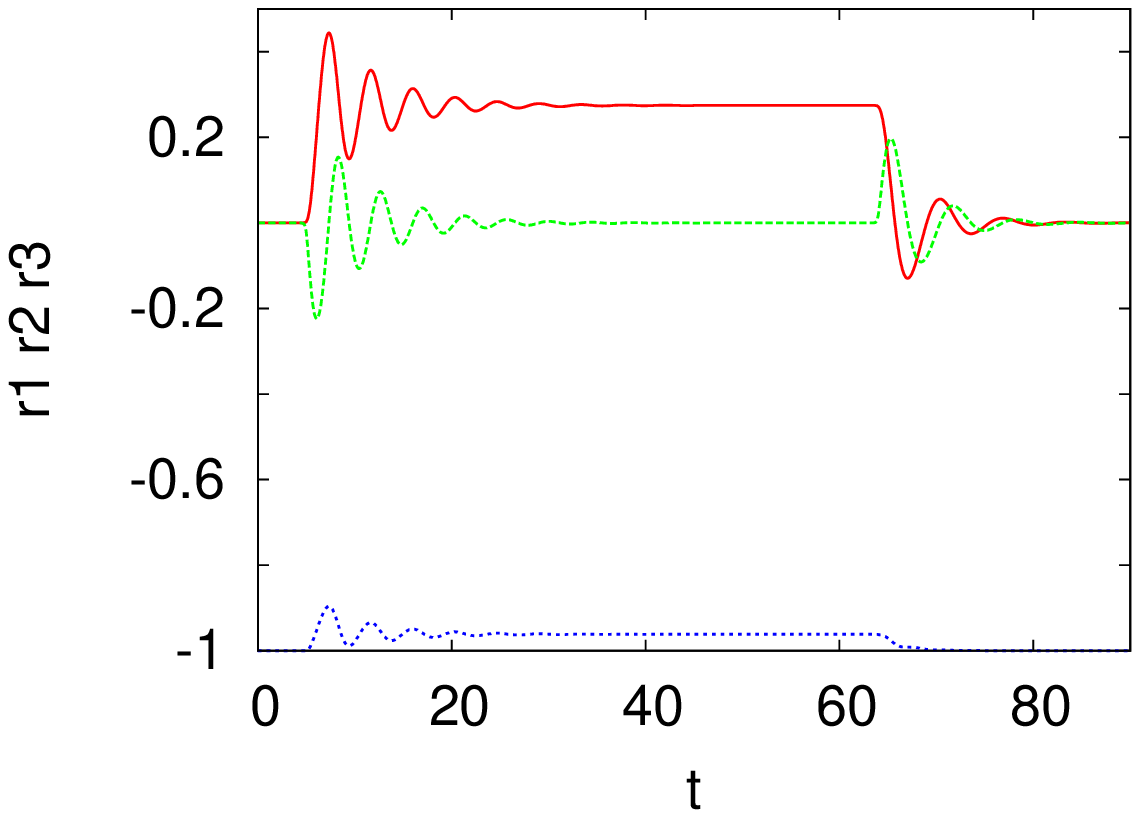,height=5 cm,width=7 cm,angle=0}
\epsfig{file=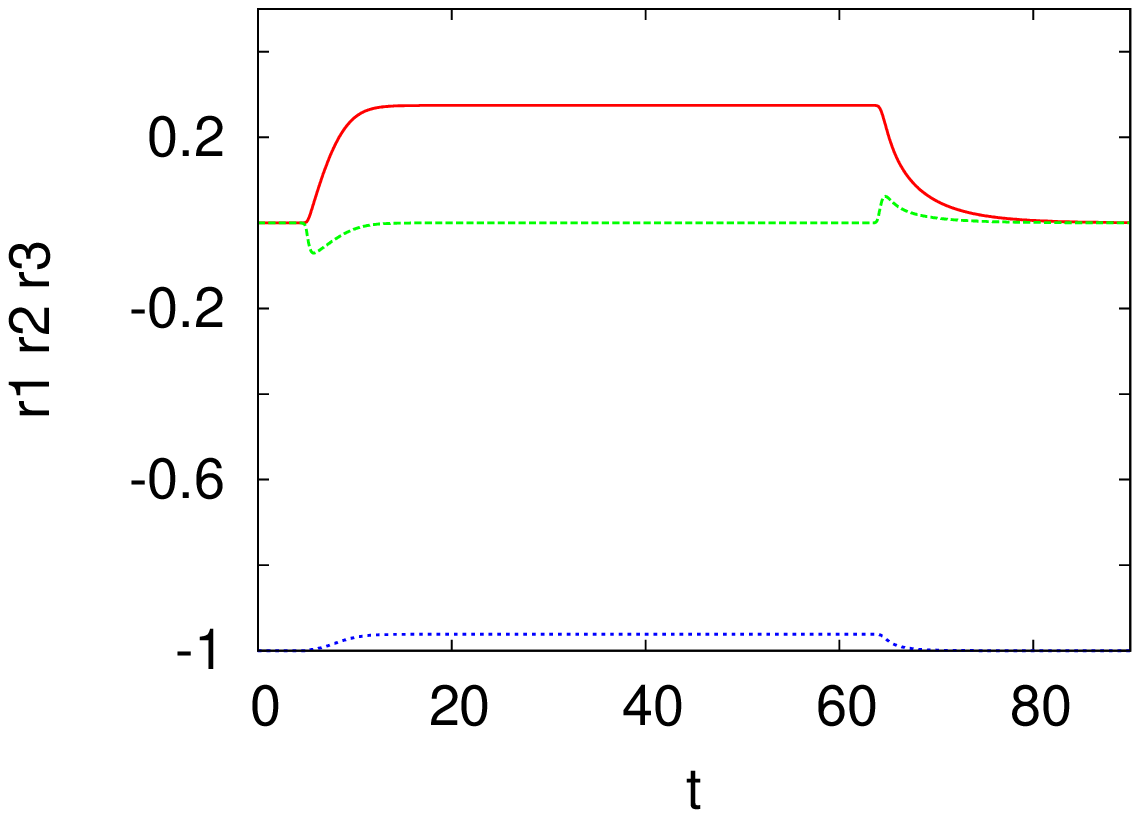,height=5 cm,width=7 cm,angle=0}
}
\caption{Time evolution of the components of the Bloch vector
$(r_1,r_2,r_3)$ respectively in continuous line, long dash and short dash
(red, green and black online) for two values of the coupling
parameter 
$\alpha=1$ for the left plot and $\alpha=10$ for the right plot.
The amplitude of the electric field is $e_0=0.4$. All the other parameters
are as in  Fig. \ref{f4}.}
\label{f6}
\end{figure}

When $\alpha$ is large, the fixed point does not change but the
eigenvalue of the Jacobian is now real. We then get no oscillations
as the system reaches the equilibrium as shown in the right
panel of Fig. \ref{f6} for which $e_0=0.4, \mu =1$ and $\alpha=10$.
Reducing $\alpha$ to 1 restores the oscillations of frequency
$\omega^*$ as shown in the left panel of Fig. \ref{f6}.

\section{Spectroscopic analysis of the film}

We now assume a constant field $e_0$ applied to the film together with
a small harmonic wave. It is then possible to examine
how these waves of different frequencies get scattered by the film.
Using this spectroscopic analysis of the film, we will see that
one can recover the atomic dipolar parameter $\mu$ and the coupling
parameter $\alpha$.

For a constant $e_0$, the system (\ref{eq27},\ref{eqm2}) becomes
\begin{eqnarray}
 \delta e_{\tau\tau}-\delta e_{\zeta\zeta} &=&\alpha \delta_D \left( x\right)\delta r_{2,\tau},  \label{eq1} \\
 \delta r_{1,\tau}&=-&(1+\mu e_{0}) \delta r_2,\\
 \delta r_{2,\tau}&=&(1+\mu e_{0})\delta r_1+e_0 \delta r_3+(\mu r_1^{\ast}+r_3^{\ast})\delta e,\\
 \delta r_{3,\tau}&=&-e_0 \delta r_2.
\end{eqnarray}
where we temporarily use $\delta_D$ to define the Dirac delta
function.

We separate time and space
\be \delta e=E(\zeta)e^{-i\omega \tau}, ~~~~\delta
r_{1,2,3}=R_{1,2,3}e^{-i\omega \tau} ,\ee
and obtain the system
\begin{eqnarray}
 E_{\zeta\zeta}+\omega^2 E&=&i\omega \alpha \delta \left( \zeta\right)R_{2},  \label{eq2} \\
 i \omega R_{1}&=&(1+\mu e_{0}) R_2,\\
 -i\omega R_{2}&=&(1+\mu e_{0}) R_1+e_0 R_3+(\mu r_1^{\ast}+r_3^{\ast})E,\\
 i \omega R_{3}&=&e_0 R_2.
\end{eqnarray}
From the second and fourth equations we get
\be R_1=-i\frac{(1+\mu e_{0}) }{\omega} R_2,~~~~ \ R_3=-i\frac{e_0
}{\omega} R_2 .\ee 
Plugging these expression into (\ref {eq2}) results in  
\be R_2=i\frac{(\mu r_1^{\ast}+r_3^{\ast})\omega
E}{\omega^2-\Omega^2}, \ee where $\Omega$ is given
by (\ref{big_omega}). 
The substitution of the previous equation into (\ref {eq1}) leads to
the non standard eigenvalue problem
\begin{eqnarray}
 E_{xx}+\omega^2 \left [1+\frac{\alpha(\mu r_1^{\ast}+r_3^{\ast})\delta(x)}{\omega^2-\Omega^2} \right]E=0.
\end{eqnarray}

As usual we assume a scattering experiment so that the field
is given by
\begin{eqnarray}\label{eq3}
E&=&\left\{ \begin{array}{l l}
A e^{i \omega x}+B e^{-i \omega x},  & x<0,\\
C e^{i \omega x} , & x>0.
\end{array}\right . 
\end{eqnarray}
At the film $x=0$ the field is continuous and its gradient
satisfies the jump condition
\begin{eqnarray}\label{eq4}
\left\{ \begin{array}{l l}
E(0+)=E(0-)=E(0)\\
E_{x}(0+)-E_{x}(0-)+\beta E(0)=0.
\end{array}\right . 
\end{eqnarray} where $$\beta=\frac{\alpha(\mu
r_1^{\ast}+r_3^{\ast})\omega^2}{\omega^2-\Omega^2}$$
Writing the two conditions (\ref{eq4}) using the
left and right fields results in
\be
\label{eq5}
\left\{ \begin{array}{l l}
A+B=C\\
i\omega(C-A+B)+\beta C=0.
\end{array}\right . 
\ee
The solution is
$$B=\frac{i\beta}{2\omega}\frac{1}{1+\frac{i\beta}{2\omega}}A, ~~~~
C=\frac{1}{1+\frac{i\beta}{2\omega}}A$$
The fixed point is such that 
$(1+\mu e_0)r_1^{\ast}+e_0r_3^{\ast}=0$, so that $\mu
r_1^{\ast}+r_3^{\ast}=-\frac{r_1^{\ast}}{e_0}$, 
so the reflection and transmission coefficients are
\be R={B\over
A}=-\frac{i\alpha\omega(r_1^{\ast}/e_0)}{2(\omega^2-\Omega^2)+i\alpha\omega(r_1^{\ast}/e_0)}
\label{eq6} \ee
 \be T={C\over
A}=\frac{(\omega^2-\Omega^2)}{(\omega^2-\Omega^2)+i\alpha\omega(r_1^{\ast}/2e_0)}
\label{eq7}\ee
where the reflection and transmission coefficients satisfy
$|R|^2+|T|^2=1$
From (\ref{fix_point}) we have 
\be r_1^{\ast}/e_0=\pm 1/\Omega, \ee
so that the transmission coefficient is
\be T=\frac{2\Omega (\omega^2-\Omega^2)
}{2\Omega(\omega^2-\Omega^2)   \pm  i\alpha \omega}.\ee
The modulus squared of the transmission and reflection coefficients
are then
\be\label {trans2} |T|^2=\frac{4\Omega^2 (\omega^2-\Omega^2)^2
}{4\Omega^2(\omega^2-\Omega^2)^2 + \alpha^2 \omega^2},\ee
\be\label {ref2} |R|^2=\frac{\alpha^2 \omega^2}
{4\Omega^2(\omega^2-\Omega^2)^2 + \alpha^2 \omega^2},\ee

\begin{figure}
\centerline{
\epsfig{file=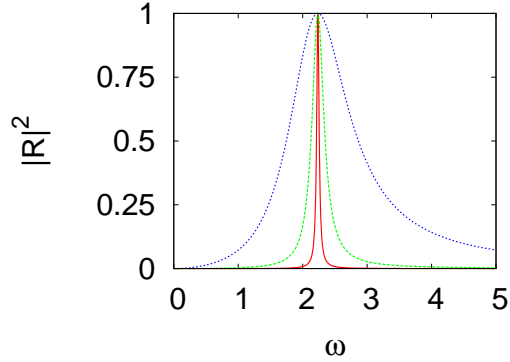,width=0.6\linewidth,angle=0}
}
\caption{Square of the modulus $|R|^2$ of the reflection
coefficient as a function of the frequency $\omega$ for three different
values of the coupling coefficient $\alpha=0.2$ (continuous line, red online)
, 1 (long dashed line, green online)  and 5 (short dashed line, blue online).
The parameters are $\mu=1, e_0=1$.
}
\label{f7}
\end{figure}

Let us examine how $|R|^2$ depends on the different parameters.
It attains its maximum 1 for $\omega = \Omega$ and decays
at infinity as ${1 \over \omega^2}$. The half-width of the
resonance,
such that $|R|^2(\omega_h)=1/2$ can be easily obtained.
Solving the quadratic equation for $\omega_h^2$ we get
\be\label{halfwidth}
\omega_h^2 = \Omega^2 \pm {\alpha \over 2} 
\sqrt {1 + {\alpha^2 \over 16 \Omega^4}} +{\alpha^2 \over 8 \Omega^2}
\approx \Omega^2 \pm {\alpha \over 2},\ee
for large $\Omega$.
Therefore the half-width is
\be\label{omegah}
\omega_h \approx \Omega \pm {\alpha \over 4 \Omega}.\ee
Fig. \ref{f7} shows the square of the modulus $|R|^2$ for
a fixed $\Omega$ ($e_0=1,~ \mu=1$). As expected the half-width
is proportional to $\alpha$. Notice how the resonance becomes
asymmetric for large $\alpha$ indicating that all higher order
terms in (\ref{halfwidth}) should be considered.

\begin{figure}
\centerline{
\epsfig{file=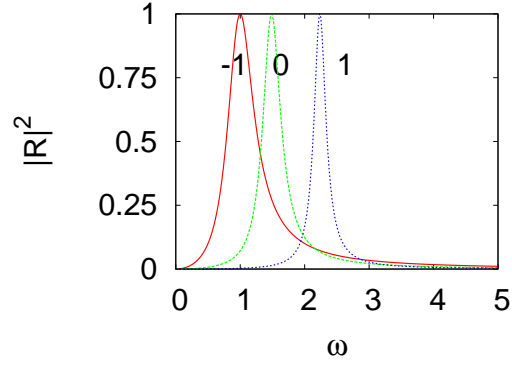,width=0.6\linewidth,angle=0}
}
\caption{Square of the modulus $|R|^2$ of the reflection
coefficient as a function of the frequency $\omega$ for three different
values of the dipole parameter $\mu=-1,~0.1$ and 1 as indicated in the
plot. 
The parameters are $e_0=1,~\alpha=1$.
}
\label{f8}
\end{figure}

When $\mu$ is varied $\Omega$ varies so $|R|^2$ will be shifted.
Fig. \ref{f8} shows $|R|^2$ for $\mu=-1,0$ and 1. As expected
for $\mu<0$ (resp. $\mu>0$) the resonance is shifted towards low 
(resp. high) frequencies. For $\mu=-1$, the higher order terms
in (\ref{halfwidth}) should be taken into account and the
resonance is not symmetric.  When $\mu=1$, the higher order terms
can dropped and the resonance curve becomes symmetric.

\begin{figure}
\centerline{
\epsfig{file=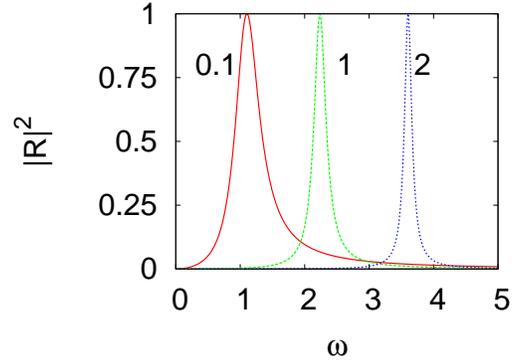,width=0.6\linewidth,angle=0}
}
\caption{Square of the modulus $|R|^2$ of the reflection
coefficient as a function of the frequency $\omega$ for three different
values of the field amplitude $e_0=0.1,~1$ and 2. 
The parameters are $\mu=1,~\alpha=1$.
}
\label{f9}
\end{figure}

The amplitude of the incoming pulse $e_0$ will also change
the resonance by shifting $\Omega$. Fig. \ref{f9} shows
$|R|^2$ for $e_0=0.1, 1$ and 2. As expected, for $e_0=0.1$
the resonance is asymmetric. It becomes symmetric for $e_0 \ge 1$.

The reflection curve $R(\omega)$ allows to estimate the
atomic parameter $\mu$ and coupling parameter $\alpha$
through the formulas (\ref{big_omega}) and (\ref{halfwidth}).
The position of the resonance gives $\Omega$ and from there
one can compute $\mu$. We have
\be\label{muofe0}
\mu = -{1 \over e_0} \pm {\sqrt{\Omega^2 -e_0^2}\over e_0}. \ee
The coupling parameter $\alpha$ can then be obtained
from (\ref{halfwidth}).

\section{Conclusion}

We consider the electromagnetic pulse propagation thought
the thin film containing the two-level atoms with taking account
both non-diagonal and diagonal matrix elements of the operator of
the dipole transition between resonant energy levels. 
The exact solution of the wave equation allows to derive the
modified Bloch equations. Thus the problem of propagating extremely
short (one- or few-cycle) pulses through a thin film is reduced
to the analysis of a system of nonlinear ordinary differential equations
on the Bloch sphere.

In the presence of a constant background field the equilibrium state
of the system film/field is changed. These new equilibrium states 
are the fixed points of the modified Bloch equations. They depend
on the field amplitude, the difference of diagonal elements of the
operator of the dipole transition (dipole parameter) and the coupling
constant. The stability
analysis of these fixed points indicates which states will be
attained by the system. For the stable states
the population difference is
negative or positive depending on the sign of the dipole parameter.
When the film is illuminated by an electromagnetic field it 
reaches the new stable state with a typical relaxation time which
depends on the field amplitude, the dipole parameter and the coupling
constant. Using our estimate, experimentalists could measure
the dipole parameter.

In the last part of the article we considered that in addition
to a constant background, the thin film is irradiated by a small
harmonic field of frequency $\omega$. This spectroscopy analysis yields the
reflection and transmission coefficients of the film as a
function of $\omega$. As expected the film is completely opaque i.e.
its reflection coefficient is equal to 1 for the field modified
transition frequency $\Omega$ given by (\ref{big_omega}). We found that
the width of the resonant curve is proportional to the
coupling constant. The position of the resonance depends on the
dipole parameter and the ground field.

The modern progress in nano-technology allows to 
produce thin films of different
features and the investigation of such features is attractive.
Our study shows that the fast electromagnetic response of a thin
film could be used in experiments to measure intrinsic  parameters
of generalized atoms (quantum dots, meta-atoms, molecules \dots)
and their coupling parameter to  the field.

{\bf Acknowledgments}

One of the authors (A.I.M.) is grateful to the
\textit{Laboratoire de Math\'ematiques, INSA de Rouen} for
hospitality and support. Elena Kazantseva thanks the 
Region Haute-Normandie for
a Post-doctoral grant. Jean-Guy Caputo 
thanks the Centre de Ressources
Informatiques de Haute-Normandie for access to computing ressources.
The authors express their gratitude to
S.O. Elyutin for enlightening discussions.

\end{document}